\documentclass[12pt,preprint]{aastex}

\shorttitle{Be Star Disks Constrained by Interferometry}
\shortauthors{Jones et al.}

\begin{document} 

\title{A Parameter Study of Classical Be Star Disk Models Constrained by 
Optical Interferometry}

\author{C. E. Jones\altaffilmark{1}, C. Tycner\altaffilmark{2}, 
T. A. A. Sigut\altaffilmark{1},
J. A. Benson\altaffilmark{3}, 
D. J. Hutter\altaffilmark{3}}

\altaffiltext{1}{Department of Physics and Astronomy, The University 
of Western Ontario, London, Ontario, N6A 3K7, Canada}
\altaffiltext{2}{Department of Physics, Central Michigan University, Mt. Pleasant, MI 48859, US}
\altaffiltext{3}{US Naval Observatory, Flagstaff Station, 
10391 W. Naval Observatory Rd., Flagstaff, AZ, 86001-8521, USA}

\slugcomment{ Accepted by ApJ}

\begin{abstract}

We have computed theoretical models of circumstellar disks for the
classical Be stars $\kappa$ Dra, $\beta$ Psc, and $\upsilon$ Cyg.
Models were constructed using a non-LTE radiative transfer code developed
by \citet{sig07} which incorporates a number of improvements over previous
treatments of the disk thermal structure, including a realistic chemical
composition.  Our models are constrained by direct comparison with
long baseline optical interferometric observations of the H$\alpha$ emitting 
regions
and by contemporaneous H$\alpha$ line profiles.
Detailed comparisons of our predictions with H$\alpha$
interferometry and spectroscopy place very tight constraints on the
density distributions for these circumstellar disks. 


\end{abstract}

\keywords{stars: circumstellar matter 
-- stars: emission line, Be 
-- stars: individual ($\kappa$ Dra, $\beta$ Psc, $\upsilon$ Cyg)
-- techniques: interferometric}


\section{Introduction}

Classical Be (or B-emission) stars are rapidly rotating, hot stars with
optical spectra that show hydrogen emission lines, and, frequently,
emission lines from singly ionized metals. It has been recognized
since the days of \citet{str31} that recombination in a flattened
disk of circumstellar gas can reproduce the basic features of the
spectroscopically observed H$\alpha$ line profiles of Be stars. However,
it is now generally accepted that this model is too simplistic.
The detailed mechanism(s) that creates and maintains Be star disks
remains unclear \citep{por03, owo03}, although rapid rotation of the
central B star certainly plays a role \citep{tow04}.  In addition to
this lack of a successful dynamical model, the evolutionary status of Be
stars and related stars (such as B[e] stars) is not well understood, and
possible evolutionary connections between these groups are ambiguous. These
problems are compounded by the additional uncertainty in their rotation
rates \citep{cra05, tow04, owo03} and lack of understanding of how rotation and
evolution interplay. The suggestion by \citet{tow04} that the rotation
rates of Be stars may be systematically underestimated, combined with
new interferometric data, has rekindled interest in finding suitable
dynamical models for these stars.

Until recently, the bulk of the observational evidence for the presence
of disks surrounding Be stars has been spectroscopic and polarimetric
in nature. The spectroscopic evidence includes the profiles of the
emission lines observed in the optical and IR spectra, as well as 
continuum excesses at infrared wavelengths.  The net linear polarization
observed in Be stars arises via electron scattering in the non-spherical
distribution of gas, as first suggested by \citet{coy69}.  However,
interferometric observations that spatially resolve the disks in
several wavelength regimes are becoming available, \citep[see
for example][]{che05, mca05, tyc06}.  Interferometric observations,
in concert with other observables, allow the disk physical properties
to be determined with greater accuracy and this may contribute to
the development of more successful dynamical models.

\citet{tyc06} collected long-baseline, optical, interferometric
observations of the classical Be stars $\gamma\;$Cas and $\phi\;$Per
at the Navy Prototype Optical Interferometer (NPOI) and interpreted
these observations using simple models of the intensity distribution
on the sky, such as uniform disks, rings, or disks with a Gaussian
intensity distribution.  In this paper, we extend the
work of \citet{tyc06} by computing detailed theoretical models for
the intensity distribution on the sky produced by the central B star
and disk of a Be system. We then compare these models to observations
collected at the NPOI. 

The theoretical models are computed with the {\sc
bedisk} code developed by \citet{sig07}. Assuming a 
density distribution of the disk gas, the code can compute the temperatures in 
the
disk given the energy input from the
central stars' photoionizing radiation field, by enforcing radiative 
equilibrium. As the disk density model
contains several adjustable parameters, values for these parameters can
then be extracted from the match to observations.  We present new observations
and detailed models for the Be stars $\kappa\;$Dra, $\beta\;$Psc, and 
$\upsilon\;$Cyg. We also use near-contemporaneous observations of the
H$\alpha$ spectral lines for these stars as additional constraints on the
density model. As we shall demonstrate, the additional spectroscopic
constraint of detailed line profiles is often an
important ingredient in selecting among models consistent with the
interferometric observations. The overall goal of this study is to place
tight constraints on physical conditions in the circumstellar regions
of these Be stars.

\section{Theory}

The theoretical disk models presented in this investigation were
constructed using the {\sc bedisk} code developed by \citet{sig07}.
A brief overview of the code and assumptions relevant to this
work are presented below. The reader is referred to \citet{sig07} for
more details.

The circumstellar disk is assumed to be axi-symmetric about the star's
rotation axis and symmetric about the mid-plane of the disk. In this
case, cylindrical geometry is appropriate and we use $R$ for the radial
distance from the star's rotation axis and $Z$ for the height above the
equatorial plane.  The radial density distribution in the equatorial
plane is assumed to be given by an $R^{-n}$ power-law following the works
of \citet{wat86,cot87, wat87}. At each radial distance from the rotation
axis, the vertical structure of the disk is determined by the requirement
of hydrostatic equilibrium in the $Z$ direction which balances the $Z$
component of the star's gravitational acceleration with the gradient of the 
gas pressure.

Typical values for the radial power-law index $n$ for theoretical
models of Be stars usually fall in the range of 2 to 3.5 based on
models that fit the IR continuum \citep{wat86}. Models with the highest
values of $n$ have a faster decrease in the density distribution with
increasing $R$. The density distribution within the disk is determined
by an assumed value of the density at the stellar surface in the equatorial 
plane, $\rho_\circ$,
by the power-law index, $n$, and by the assumption of pressure
support perpendicular to the equatorial plane. Since $\rho_\circ$
and $n$ significantly affect the theoretical predictions, we conduct a
2-dimensional parameter search by varying these input parameters over
all reasonable values for comparison with interferometric and 
spectroscopic observations.

Given this density model, the {\sc bedisk} code is used to find the
temperature structure of the disk by enforcing radiative equilibrium.
The microscopic heating and cooling rates implied by a gas with a solar
chemical composition are computed and the temperature that balances
heating and cooling is found. To compute the atomic level populations
required by this procedure, the statistical equilibrium equations are
solved for each chemical element included, accounting for the bound-bound
and bound-free collisional and radiative processes that set the rates
in and out of each atomic level. 

To compute the hydrogen line profiles, we solved the transfer equation
along a series of rays through the star plus disk system as viewed at an
angle $i$ relative to the line-of-sight (where $i=0^o$ is pole-on,
and $i=90^o$ is equator-on).  The disk is assumed to be in pure
Keplerian
rotation, and the equatorial rotational velocity of the star was chosen
so that the measured $v\sin i$ of the star was recovered for the adopted
value of $i$.  For rays terminating on the stellar surface, we adopted
a photospheric H$\alpha$ line profile computed using the {\sc synthe}
code and the Stark Broadening routines of \citet{bar03}; these profiles
are based on the hydrogen populations computed for the appropriate LTE,
line-blanketed model atmosphere adopted from \citet{kur93}. 
For rays passing through the disk, the equation of radiative transfer
was solved along the ray using the short-characteristics method
of \citet{ols87}. To compute the H$\alpha$ opacity and
emissivity, the hydrogen level populations computed for the thermal
solution were used. Here, hydrogen was represented by a 15 level atom
and all implied radiative and collisional bound-bound and bound-free
processes from and between the 15 levels were included. The radiative
bound-bound rates were treated by the escape probability approximation
in which the net radiative bracket for the line \citep{mih78} is
replaced by a single-flight escape probability. 
While the escape
probability approximation can provide a reasonable description of the
line thermalization, the use of escape probabilities is an important
approximation. The {\sc bedisk} code assumes static escape probabilities
based on the complete redistribution over a Doppler profile for all
lines \citep[for a justification of this procedure see][]{sig07}.
If the functional form of the escape probability function is changed,
for example, to a simple $\tau^{-1}$ dependence on optical depth for
$\tau\gg1$, the predicted equivalent width of H$\alpha$ can change by
$\sim\,2$ \AA $\,$ (due to the change in the thermal structure of
the disk and in the hydrogen level populations). Thus in matching to the
observed H$\alpha$ profiles, the largest uncertainty is likely from the
uncertainty in the theoretical profile and not from the observational
uncertainties (see Section 3 for an estimate of the observational 
uncertainties).

Our approach allows 
the full dependence of the H$\alpha$ line emissivity and opacity on
the physical conditions throughout the disk to be retained. The Stark
broadening routines of \citet{bar03} were also used to compute the
local H$\alpha$ line profile throughout the disk.  This is important as
these routines can handle high disk densities for which the assumption
of a simple Gaussian or Voigt profile for the H$\alpha$ line would be
invalid. The final H$\alpha$ line profile from the unresolved system
was obtained by summing over the rays weighted by their projected area
on the sky. The H$\alpha$ line profile was computed over a wavelength bin of 
$\pm20$\AA\ from
line centre at each area element on the sky. To obtain the final H$\alpha$ 
line profile to
compare with spectroscopic observations, the profile was convolved
with a Gaussian of FWHM of $0.656\,$\AA\ to bring the resolving power of the
computed profile down to $10^4$ to match the observations.  The predicted 
interferometric 
visibilities also followed
from the same numerical model. The monochromatic H$\alpha$ image of the system 
projected on the plane of the sky was computed by integrating
the stellar continuum and H$\alpha$ line flux over a 150 \AA $\,$ spectral window
centered at
H$\alpha$ to represent the interferometric spectral channels (see Section 3).
Thus the calculation of the
H$\alpha$ profiles and the H$\alpha$ interferometric visibilities
used the same model calculations. 

\section{Observations}

We obtained long-baseline interferometric and spectroscopic
observations of three Be stars, $\kappa$~Dra, $\beta$~Psc, and
$\upsilon$~Cyg.  The interferometric observations were acquired using the
Navy Prototype Optical Interferometer~(NPOI) utilizing baselines
ranging from 18.9 to 64.4~m in length.  The NPOI is described in
detail in \citet{Armstrong98} and the technique used to extract
interferometric observables from the spectral channel containing the
H$\alpha$ emission line has been discussed by \citet{Tycner03}.
Table~\ref{obs_table} shows all the interferometric observations
obtained for the three stars and the number of data points obtained on
each night in the spectral channel containing the H$\alpha$ line.  The
interferometric observations of all stars were obtained over a period
of less than a month, with $\beta$~Psc and $\upsilon$~Cyg
observed during 2005, and $\kappa$~Dra 
observed during 2006. 

To complement our interferometric observations we have also obtained
near contemporaneous spectroscopic observations of all three targets.
The spectroscopic observations were obtained using an Echelle
spectrograph at the Lowell Observatory's John. S. Hall telescope and
covered the spectral region around the H$\alpha$ line.  The properties
of the spectrograph and the description of the reduction methods have
been described elsewhere~\citep[see \S~3.2 in][and references
therein]{tyc06}.  
The continuum signal-to-noise ratio (SNR) near H$\alpha$  
is more than 200 in the spectra of all three stars.
However, the equivalent width and the shape of the line is 
determined with  
respect to the normalized continuum level.  Because the  
line profile must be normalized by a smoothly varying function  
that is fit to the continuum (or where one thinks the continuum is  
located), the choice of this normalizing function affects the shape  
of the H$\alpha$ line.
In addition, the presence of telluric lines, as well as their temporal  
variability from night to night (and season to season) can affect how  
well the continuum level is determined; we expect this to be the  
dominant source of uncertainty associated with the shape of the  
observed H$\alpha$ line.  We have estimated the magnitude of this effect by  
analyzing a set of reduced spectra all obtained from the same raw  
spectrum, but all obtained with slightly different quadratic  
functions fitted to slightly different continuum regions.   
The variations in the resulting normalized continuum were at the 2 to  
3\% level, and to be conservative we adopt a 3\% uncertainty for our  
continuum level determination and, in turn, the shape and equivalent width   
measure of the observed H$\alpha$ emission line.

For $\beta$~Psc and $\upsilon$~Cyg the spectra were
obtained on 2005~Sep~16, and for $\kappa$~Dra the spectrum was
obtained on 2006~Mar~17, and therefore in all three cases the
spectroscopic observations overlapped the interferometric observing
runs~(Table~\ref{obs_table}). The equivalent widths of the observed 
H$\alpha$
spectra lines are -15.3 \AA, -24.8 \AA, and -22.2 \AA $\,$ for 
$\beta$~Psc, $\upsilon$~Cyg, 
and $\kappa$~Dra, respectively.  (We adopt the usual convention that a negative 
in 
the quoted value of the equivalent width indicates a net line flux 
above the continuum.)

\begin{table}
\begin{center}
\caption[]{\sc \small Observing Log for Interferometric Observations \label{obs_table}}
\begin{tabular}{lccc} \hline\hline
\hspace{2cm} UT Date \hspace{2cm}      &    $\kappa$ Dra  &  $\beta$ Psc   &  $\upsilon$ Cyg  \\
                                       & (\# of points)    &  (\# of points) &  (\# of points)   \\\hline
2005 Aug 28                   \dotfill &   \ldots         &        12      &        8         \\
2005 Aug 29                   \dotfill &   \ldots         &        16      &       12         \\
2005 Aug 30                   \dotfill &   \ldots         &        20      &       20         \\
2005 Aug 31                   \dotfill &   \ldots         &        20      &       20         \\
2005 Sep 1                   \dotfill &   \ldots         &         20      &       20         \\
2005 Sep 5                   \dotfill &   \ldots         &          8      &        4         \\
2005 Sep 13                  \dotfill &   \ldots         &         16      &       16         \\
2005 Sep 14                  \dotfill &   \ldots         &         23      &        20         \\
2005 Sep 15                   \dotfill &   \ldots         &        20      &        20         \\
2005 Sep 16                   \dotfill &   \ldots         &        21      &        24         \\
2005 Sep 17                   \dotfill &   \ldots         &         8      &        9         \\
2005 Sep 26                   \dotfill &   \ldots         &        16      &        28         \\
2006 Feb 25               \dotfill &       30      &     \ldots      &    \ldots  \\
2006 Feb 26               \dotfill &        6      &     \ldots      &    \ldots  \\
2006 Mar 5                \dotfill &       52      &     \ldots      &    \ldots  \\
2006 Mar 6                \dotfill &       24      &     \ldots      &    \ldots  \\
2006 Mar 14               \dotfill &       36      &     \ldots      &    \ldots  \\
2006 Mar 16               \dotfill &       76      &     \ldots      &    \ldots  \\
2006 Mar 19               \dotfill &       52      &     \ldots      &    \ldots  \\
Total:                             &      276      &     200         &     201   \\
\hline
\end{tabular}
\end{center}
\end{table} 

\section{Results}

There is considerable uncertainty in the assigned spectral types
for Be stars due to rapid rotation and spectral variability.  The effects
of gravity darkening, and the potential obscuration of portions of the 
stellar surface
by the disk, further compound the uncertainty in
classification and the corresponding stellar parameters. Since the stellar
parameters fix the photoionizing radiation field that is assumed to be
the sole source of energy input into the  circumstellar disk, realistic
stellar parameters are essential to the construction of reasonable disk
models.  For each of the stars in this study, we searched the literature
in order to find the most reasonable stellar parameters available.
See Table~\ref{stellar_param} for the adopted stellar parameters for
$\kappa$ Dra, $\beta$ Psc, and $\upsilon$ Cyg.  The rationale for adopting
the particular stellar parameters is explained and compared with other
published values in each subsection for a given star.

\begin{table}
\begin{center}
\caption{Adopted stellar parameters \label{stellar_param}}
\begin{tabular}{lccccccccr} 
\tableline\tableline
Star& HR& $\log g$&Spectral& Stellar Radius & Stellar Mass & $T_{eff}$ &$v\sin i$& Reference\\
Name&Number &&Type& ${\rm R}_\odot$ & ${\rm M}_\odot$ & K &km/s\\
\tableline
$\kappa$ Dra  & HR4787 &3.5& B6IIIpe &6.4 &4.8   & 14000 & 170       &1 \\
$\beta$ Psc   &HR8773  &4.0& B6Ve    &3.6 &4.7   & 15500 & 90$\pm$15 &  2 \\
$\upsilon$ Cyg&HR8146  &4.0& B2Vne   &4.7 &6.8   & 19800 & 173$\pm$10&3 \\
\tableline
\end{tabular}
\tablerefs{ (1)\citet{saa04}; (2)\citet{lev04}; (3) \citet{nei05}}
\end{center}
\end{table}

\subsection{$\kappa$ Draconis}

$\kappa$ Dra is a known spectroscopic binary and giant star
of intermediate spectral type.  It is variable on a variety of
timescales; see \citet{hir95} for a detailed discussion of these
variations.  \citet{saa04} modeled the central star of $\kappa$ Dra by
comparing a variety of spectroscopic and photometric observations to a
grid of non-LTE models. We have adopted their derived best-fit stellar
parameters for our study.  We used their parameters of $T_{eff}$,
log$g$, and mass to calculate the stellar radius.  \citet{gie07}
also used these parameters for the interpretation of $K^\prime$-band
observations of $\kappa\;$Dra collected at the CHARA Array interferometer.  
The parameters of \citet{saa04} are also in reasonable
agreement with stellar parameters listed for the spectral class of B5III
in \citet{cox00}.


\begin{figure}
\epsscale{.80}
\plotone{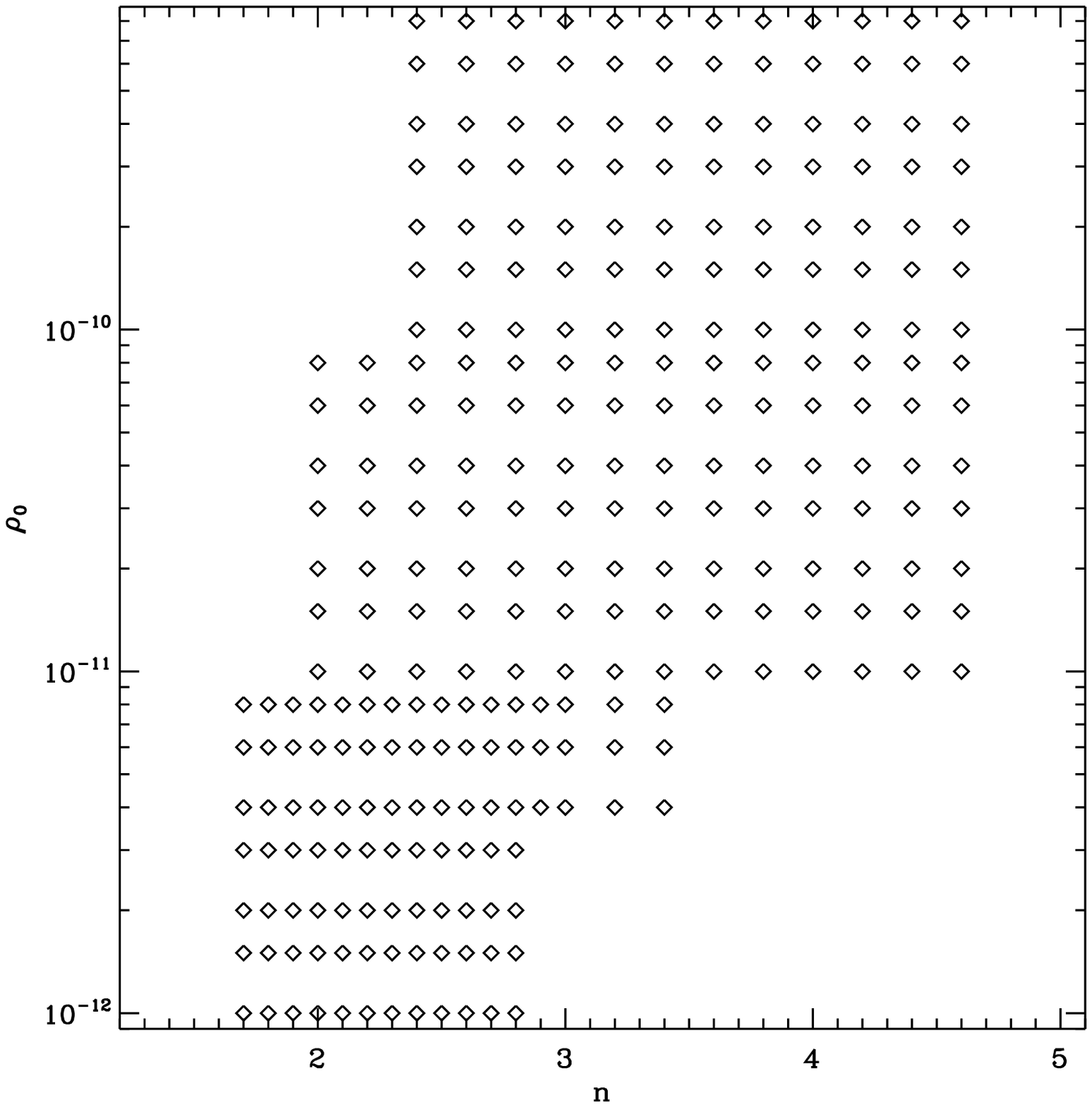}
\caption{The density parameter grid for $\kappa$ Dra shown as a function
of $\rho_o$ (g/cm$^3$) and $n$.\label{KD_grid}}
\end{figure}

\begin{figure}
\epsscale{.80}
\plotone{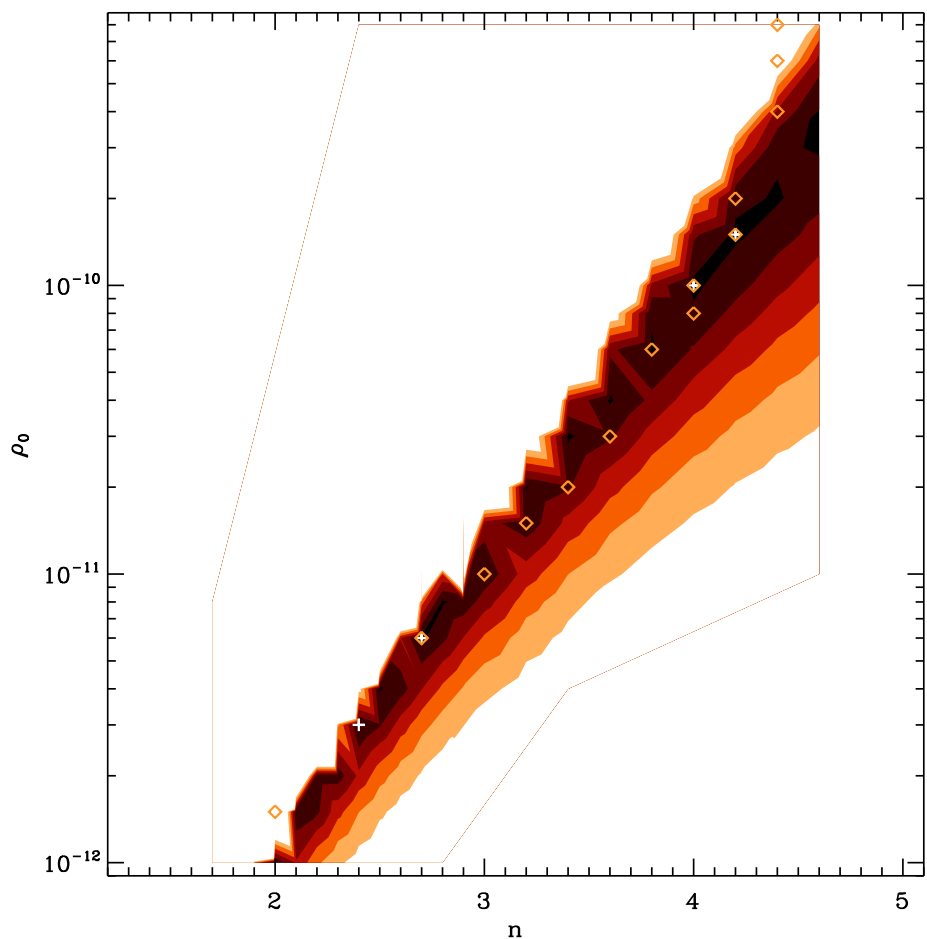}
\caption{The $\chi^2$ values for computed models as a function of $n$ 
and $\rho_o$ (g/cm$^3$) for $\kappa$ Dra. The contour plot shows the 
reduced $\chi^2$ 
values 
in steps of 1.0.  Only a region  of reduced $\chi^2 < 6$ is shown.  The 
darkest regions on the 
figure represent models with reduced $\chi^2 <2 $. The four white plus
signs correspond to models with the lowest values of reduced $\chi^2$ 
determined from interferometry. The diamonds correspond to models that
have predicted H$\alpha$ equivalent widths $\pm \, 2 \, $\AA $\,$ of the 
observation.
\label{KD_chi2}}
\end{figure}

\begin{figure}
\epsscale{.80}
\plotone{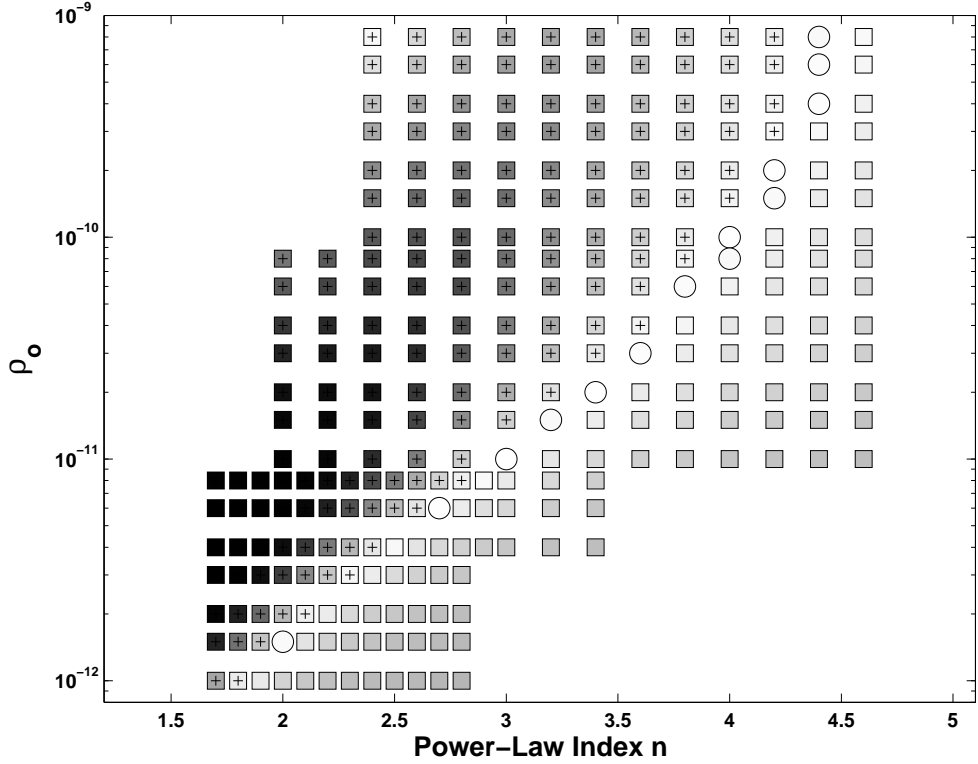}
\caption{The predicted H$\alpha$ equivalent widths as a function of
$\log \, \rho_o$, and the power-law index, $n$, for the entire grid
of models.
The lightest coloured symbols are closest to the value of the observed 
equivalent width of -22.2 \AA $\,$. The models that predict H$\alpha$
equivalent width within $\pm \, 2 \, $\AA $\,$ of the observation are represented 
by white circles. The models that correspond to predicted H$\alpha$
emission greater than the observation are indicated by the plus sign
within the symbol. \label{halpha_2d}}
\end{figure}

\begin{figure}
\epsscale{.80}
\plotone{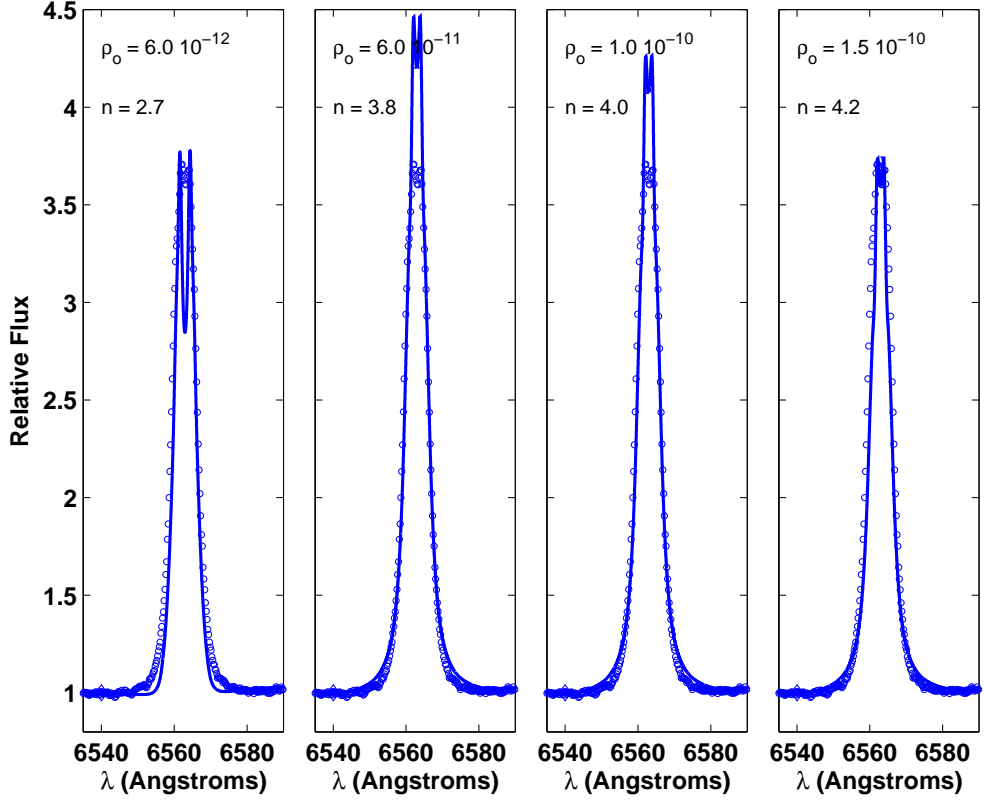}
\caption{The observed H$\alpha$ line for $\kappa$ Dra compared with 4 models 
with $n$ and $\rho_o$ as indicated in the panels. These models correspond
the best-fit from both interferometry and H$\alpha$ modeling.  The model 
profiles are the 
solid lines and the circles 
represent the observed line. All model profiles were obtained with a 
disk inclination of 
35$^o$.  The reduced $\chi^2$ values from interferometry for these models are 
1.18, 1.29, 1.18, and 1.18, from left to right respectively.
 \label{halpha_dra}}
\end{figure}

\begin{figure} 
\epsscale{.80} 
\plotone{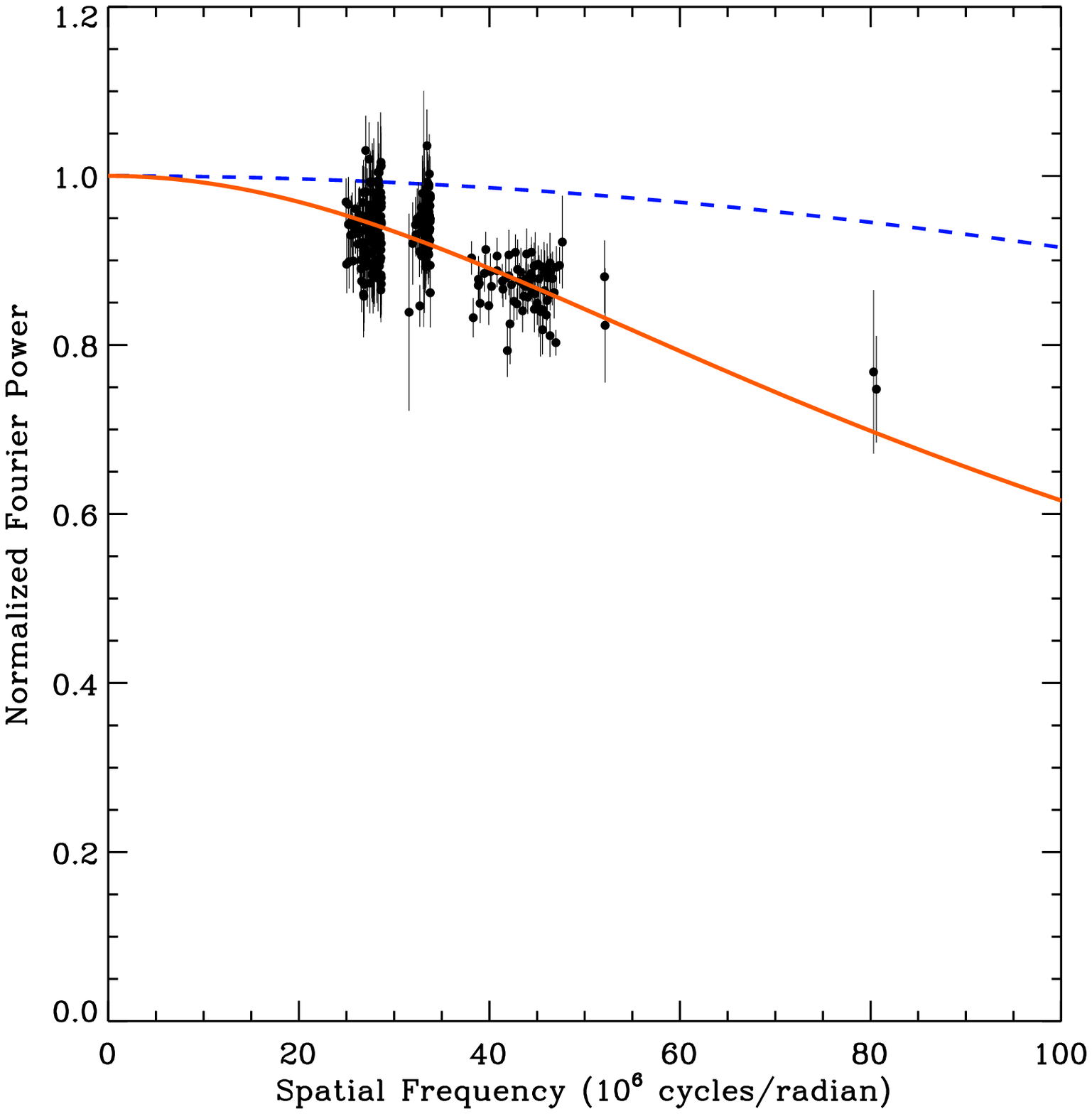} 
\caption{The interferometric data and model with $n=4.2 $, and 
$\rho_o = 1.5 \times 10^{-10}$ g/cm$^3$ for $\kappa$ Dra. The dashed line
represents the central star and the solid line is obtained by taking
a Fourier transform of a synthetic image of the preferred model.
 \label{fig_dra_fits}} 
\end{figure} 

Figure~\ref{KD_grid} shows the density parameter grid for the disk of
$\kappa\;$Dra.  This grid was selected after running a series of models
over a much wider range in $\rho_o$ and $n$ on a coarser grid. A comparison
of the observed and predicted H$\alpha$ equivalent widths allowed us
to determine a finer grid to sample for detailed comparisons with
interferometric observations.  The final density parameter grid 
consisted of 278
models. We have compared the output from each model to the 
interferometric
observations using the technique described in detail by \citet{tyc08}.
 The reduced $\chi^2$ values corresponding to these models are presented
in Figure~\ref{KD_chi2} as a function of $\rho_o$ (g/cm$^3$) and $n$. 
The darkest patches in this Figure
represent the models with reduced $\chi^2 < 2$.
The white plus signs correspond to models with the lowest values 
(ranging from 1.18 to 1.29) of the interferometric reduced $\chi^2 $.
In total, there are 25 models with reduced $\chi^2 < 2$ for this star,
therefore, it is necessary to further constrain our models in order to 
determine the best-fit model.

Figure~\ref{halpha_2d} shows the predicted H$\alpha$ equivalent widths as a 
function of $\log \, \rho_o$, and the power-law index, $n$, for the entire 
grid of models. The predicted H$\alpha$ profiles were constructed 
assuming an inclination of $i=35^o$.  We note that this value 
of $i$ is consistent with other values
determined from spectroscopy and polarization analyzes presented in
the literature.  For example, \citet{cla90}, estimates $i=23^o$ based
on the observed polarization angle, and \citet{juz91} find $i=35^o - 45^o$ 
based
on spectroscopy.  
We also experimented by varying $i$ and found that an inclination angle
for this system of $35^o$ seemed to produce profiles that matched
the observed H$\alpha$ line.  We also note that although changing the value of
$i$ by $\pm$ $10^o$ changes the shape of the H$\alpha$ profile, the effect on 
the equivalent width is $< 1$ \AA. 
The lightest coloured symbols are closest to the value of the observed 
equivalent width of -22.2 \AA. There are 14 models that predict an H$\alpha$
equivalent width within $\pm \, 2 \,$ \AA $\,$ of the observation and these 
are 
represented 
by white circles. Models which correspond to predicted H$\alpha$
equivalent widths greater than the observed emission have 
a plus sign within the corresponding symbol in Figure~\ref{halpha_2d}.

We note that this subset of 14 models
does not represent the 14 best models using interferometry alone. 
Although this
set matches the observed H$\alpha$ equivalent width within $\pm$ 2 \AA, 
using 
interferometric observations as a constraint 
this same set corresponds to reduced $\chi^2$ from 1.18 (the minimum value)
up to 16.0.  In fact, 6 of the 14 models in this set had a reduced $\chi^2$
greater than 2.0. It is clear from a comparison of  Figure~\ref{KD_chi2}
and Figure~\ref{halpha_2d} that the results from H$\alpha$ interferometry 
and H$\alpha$ line profile are complimentary but that each provides
extra information to help reduce the degeneracy in finding the best-fit
model.

Multiple models, of pairs of $n$ and $\rho_o$, that produce
H$\alpha$ profiles with equivalent widths that match the observations within
uncertainties can be found. However, not all of these are
the best-fits in terms of interferometry. 
That is, it is possible to match the equivalent width for a range of $n$
by an appropriate adjustment in $\rho_o$, but that these
combinations do not necessary have the correct density distribution as
a function of radial distance from the star.  
However,
by combining results from spectroscopy and interferometry,
we can find the best pairs of $n$ and $\rho_o$ for $\kappa$ Dra.

Next, we compared the subset of 8 models which represent the best-fits 
from both interferometry and H$\alpha$ equivalent width. A subset 
(4 models with the lowest reduced $\chi^2$ from interferometry) 
of the corresponding
H$\alpha$ profiles is shown in
Figure~\ref{halpha_dra}.  Notice the variation in the shape of the
profile as $n$ is increased from 2.7 in the left-most panel to 4.2 in
the right-most panel.  The predicted equivalent widths are -21.9,
-23.5, -22.4, and -20.3 \AA, with the corresponding $\chi^2$ values from 
interferometry of 
1.18, 1.29, 1.18, and 1.18. The model with $n=2.7$ does not have the the 
correct line profile shape
compared to the observed profile. There is too much absorption 
at line centre, and the profile is clearly too narrow in the wings.
All of the predicted profiles in Figure~\ref{halpha_dra} have an absorption
feature at line centre.  This feature is the most pronounced for low values
of $n$ and gradually decreases for models with higher values of $n$.
If we increase the inclination, this absorption feature becomes more
and more pronounced as the system is viewed closer and closer to edge
on. If we decrease $i$, so the system is viewed close to pole-on, the
absorption feature remains for the models with the lowest values of $n$.  
In addition, with
$n=2.7$, the wings of the predicted profile are not broad enough for
any reasonable values of $i$.  The panel with the model corresponding to 
$n=3.8$ and $n=4.0$ in 
Figure~\ref{halpha_dra}
shows that the predicted line is too strong and this results in an H$\alpha$
equivalent width which is too large. Models corresponding to 
$n > 4.2 $ have profiles with wings wider than the observation. 
The model that corresponds to the profile with $n = 4.2$ matches the
shape of the observed profile best and  
has a minimum reduced $\chi^2$
of 1.18 from interferometry. We prefer
this model with $n = 4.2$ and $\rho_o = 1.5 \times 10^{-10}$ g/cm$^3$
in terms of both spectroscopy and interferometry.
Figure~\ref{fig_dra_fits} shows the interferometric observations compared
with this model for $\kappa$ Dra.

We note that as previously discussed the uncertainty due to the
approximations in the computational code (see Section 2) are larger
than those in the observed H$\alpha$ profiles (see Section 3). 
We searched the literature to find the best possible stellar parameters
for the disk systems studied in this investigation
(see Section 4.1, 4.2, 4.3). However, generally for Be stars, not only are the
stellar parameters not well established but for some stars, even the 
spectral type may not be accurate.  We now wish to test the dependence 
of model parameters on our results.  We experimented by varying the
stellar parameters in turn, to see how these changes affect 
the predicted H$\alpha$ equivalent widths and the interferometric fits for
the star $\kappa$ Dra. Similar 
types of uncertainties would be expected for the other two systems presented 
in this 
paper. One might expect that the 
$T_{eff}$ of the central star, which supplies the disk with energy, may be the 
source of 
the greatest uncertainty. We changed the 
$T_{eff}$ by $\pm$ 1000 $K$ and reproduced all of the profiles. This resulted
in a change in the equivalent width of the H$\alpha$ line by a maximum of
$\sim$ 2 \AA. To assess the effect of uncertainties in the fundamental stellar
parameters combined,
we performed a Monte Carlo simulation in which new stellar parameters
were randomly realized. These parameters were used to compute a new
circumstellar disk model, corresponding to our preferred model of
$n=4.2$ and $\rho_o = 1.5 \times 10^{-10}\;\rm{g/cm^{3}}$. This procedure
was repeated to generate new H$\alpha$ profiles and equivalent
widths. Errors assumed for the stellar parameters were $\pm\,25$\%
in mass, $\pm\,25$\% in radius, and $\pm\,10.7$\% in $T_{eff}$
(corresponding to $\pm\,1500\;$K); each stellar parameter was assumed equally 
probable within its errors. Using a Gaussian distribution for the errors,
we found our results did not change significantly. The mean of H$\alpha$ 
equivalent
width was $- 20\,$\AA\ with a standard deviation of $2.65\,$\AA.

We have one star in common with the work of \citet{gie07}: $\kappa\;
$Dra. They use a disk density model which is very similar to ours and
derive two sets of parameters $(n,\rho_o)$ based on fits to K'-band
interferometric visibilities: $n=0.16\pm0.43$, $\rho_o = 2.8 \times 10^{-13}\;
\rm g\,cm^{-3}$ for
a single star model, and $n=0.67\pm0.36$, $\rho_o = 6.2 \times 10^{-13}\;\rm
g\,cm^{-3}$ for a binary model.  \citet{gie07} prefer the binary
fit to the single star model because of the peculiar solution they
obtain in the single star case; namely a very small exponent in the radial
density power law ($n=0.16$) and a best solution with $i=72^o \pm 18^o$ which
is at odds with the resolved H$\alpha$ line profile. Nevertheless, the
particular properties of the introduced binary are not constrained by
any independent observations.  Both of these solutions seem at odds with
our results. It is
well documented that the H$\alpha$ equivalent width and the infrared excess
are correlated \citep[see for example][]{van95}.
\citet{saa04} present a figure which shows the change in H$\alpha$ 
equivalent width
from the early 1970's to 2004 for $\kappa$ Dra 
(see \citet{saa04}, figure 4(d)). Although 
there is certainly scatter of data values in their figure, one can easily see 
from this 
figure that the H$\alpha$ equivalent width varies by $\sim$ 15 \AA $\,$ (from 
about - 5 to about -20 \AA ) in a period of $\sim$ 6 years.
This is a substantial change in the H$\alpha$ emission line and
we expect there would be a corresponding increase in the infrared excess. 
The interferometric observations for $\kappa$ Dra presented in \citet{gie07} 
were 
obtained in April (3 dates) and December (2 dates) of 2005. Their 
spectroscopic observations were obtained in April 2005.  Our contemporaneous 
observations were obtained in 2006. 
The difference in our results and \citet{gie07} is partially accounted 
for by the fact that we model this variable star at different times.
This further supports the requirement of contemporaneous observations. 
Other possible causes for the difference in results could be due the 
fact that the H$\alpha$ observations are sensitive to different regions
within the disk as well as the different spatial resolution of the two 
interferometers.

There are also other
differences between our calculations and \citet{gie07} that 
should
be highlighted: as noted in Section~2, the calculation of the H$\alpha$
visibilities and the H$\alpha$ line profiles in our work are obtained
from
essentially the same calculation that determined the radiative
equilibrium
temperature structure for the disk given the adopted density model. The
H$\alpha$ opacity and emissivity at each point in the disk were computed
from the local temperature, pressure, and radiative field by solving
the set of statistical equilibrium equations for a 15-level hydrogen
atom. While our fits to the observed H$\alpha$ line profiles are not
perfect, they are predicted naturally from the models and the assumption
of pure Keplerian rotation by the disk. To fit the H$\alpha$, H$\gamma$,
and Br$\alpha$ lines, \citet{gie07} found it necessary to convolve
their
profiles with a unit normalized Lorentzian to broaden the line wings;
they discuss numerous possible sources for this extra broadening. In the
current work, we convolved our line profiles only with a unit normalized
Gaussian of FWHM of $0.656\;$\AA\ to bring the resolving power of the
computed
profile to $10^4$ to match the observations. Among the reasons cited
by \citet{gie07} as to possible broad H$\alpha$ wings, they discuss
the possibility of Stark broadening for the hydrogen lines, but then
note that the densities they find are too low for Stark broadening to
be important. This is consistent with their best fit densities which
are about 1-2 orders of magnitudes smaller than those derived here.

Previous studies have shown that self-consistent disk thermal structures
are crucial in order to interpret observations correctly \citep[see
for example,][]{car06,sig07}. We computed the thermal structure for $\kappa$
Dra with our {\sc
bedisk} code using the model parameters,  
$n=0.67\pm0.36$, $\rho_o = 6.2 \times 10^{-13}\;\rm
g\,cm^{-3}$, R$_d = 67 {\rm R}_\odot$, listed in \citet{gie07}. We find 
substantial variation in 
disk temperature near the star and equatorial plane, and an overall
average disk temperature of $\sim$ 13000 K. \citet{gie07} adopt an isothermal
disk temperature of $\sim$ 9300 K for $\kappa$ Dra based on previous studies
\citep{car06}, but we note that this study as well as \citet{sig07} do not 
investigate the disk densities as low as the \citet{gie07} best-fit model. 
Finally, we note that the value, $n = 4.2$, we obtain for $\kappa$ Dra
is 
in agreement with typical values previously predicted in the literature
for Be star disks
\citep[see
for example][]{wat86,tyc08}.

\subsection{$\beta$ Pisces}

$\beta$ Psc is a B6Ve star that has long been known to exhibit emission
in the hydrogen lines \citep[see for example][]{mer25}. This star is
similar in spectral type to $\kappa$ Dra, but it is a main sequence dwarf.
We have included $\beta$ Psc in this investigation because comparing the
disk models with the interferometric observations lead to unique
density parameters, and it is useful to compare the results for this star
with $\kappa$ Dra.  We adopt the stellar parameters from \citet{lev04}
and note that these parameters are also consistent with interpolated
values in \citet{cox00}.  The interferometric observations are from
12 nights between 28 Aug. 2005 and 26 Sept.\ 2005 (see Table~\ref{obs_table}), 
and the H$\alpha$
spectroscopy for $\beta$ Psc (and $\upsilon$ Cyg) were obtained on 16
Sept.\ 2005 and therefore are definitely representative of the H$\alpha$
line during the period of the interferometric observations. We note that a
subsequent observation of the H$\alpha$ line for $\beta$ Psc on 3 
Oct.\ 2006 had a 20\% weaker equivalent width. 

The procedure for selecting the grid of models to be compared in detail
with interferometric observations was obtained by the same selection
process as done for $\kappa$ Dra. The final parameters to be searched
covered the range $n = 1.5$ to $n = 5.5$ and $\rho_o = 1.0 \times
10^{-12}$ g/cm$^3$ to   $\rho_o = 8.0 \times 10^{-9}$ g/cm$^3$. In this
range, 193 models were constructed for comparison with interferometric
observations.  The corresponding $\chi^2$ values had a minimum value of
0.83 corresponding to the model with $n=2.2$ and $\rho_o = 4.0 \times
10^{-12}$ g/cm$^3$ and a maximum value of 438 corresponding the model with
$n=1.7$ and $\rho_o = 1.0 \times 10^{-11}$ g/cm$^3$.  The $\chi^2$ values
revealed only one other model, $n=2.1$ and $\rho_o = 3.0
\times 10^{-12}$, which was statistically the same as the best-fit. For
this model $\chi^2 = 0.84$.

\begin{figure}
\epsscale{.80}
\plotone{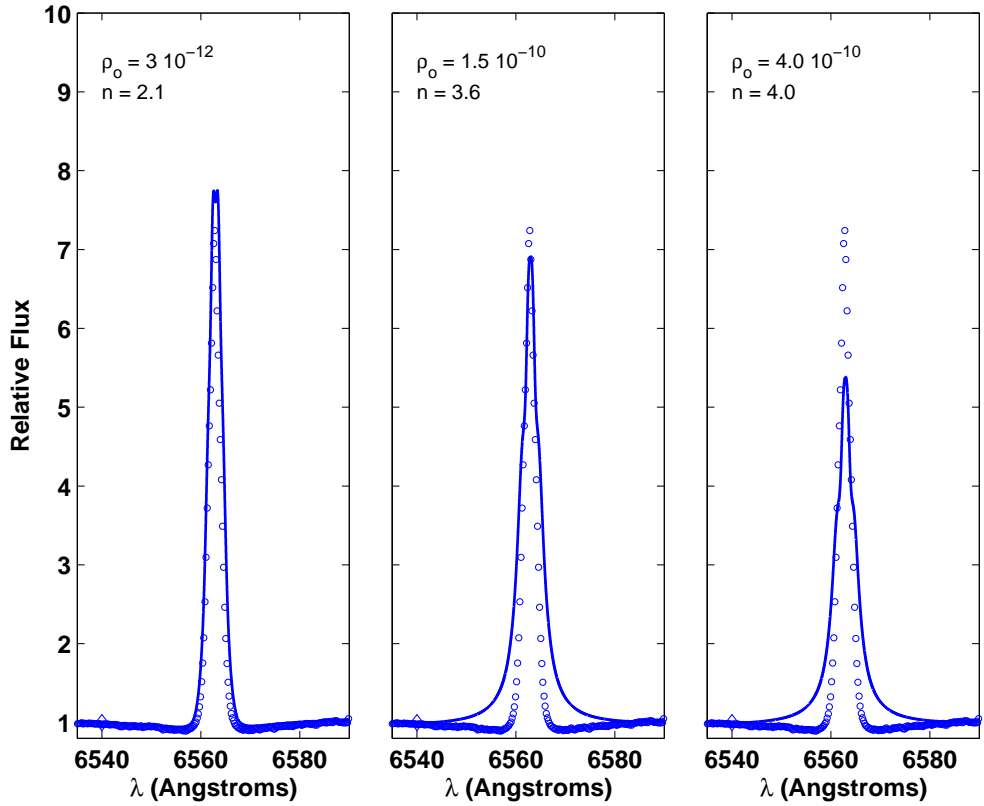}
\caption{The observed H$\alpha$ line for 
$\beta$ Psc compared with model predictions for a series of models showing the 
changes in the profile as $n$ is increased. The model profiles are the solid lines and the circles represent the observation. \label{halpha_psc}}
\end{figure}

In order to check the predicted H$\alpha$ profiles, we had to adopt a value
for $i$.  The observed sharp emission line profile (shown in
Figure~\ref{halpha_psc}) suggests a small value for $i$. However, this
star has a  $v \sin \, i$ of 90$^o$ km/s (see Table~\ref{stellar_param}),
so for very small values of $i$, the star will exceed its critical
velocity. The estimated critical velocity of a star of this spectral type
is $\sim 416$ km/s \citep{por96}. Therefore, we argue that the inclination
of this system must be at least 15$^o$, and we have adopted a value of $i
= 20^o$.

We calculated the H$\alpha$ equivalent widths for a subset ($\sim 20$
models) chosen by the best-fit $\chi^2$ values, and we considered $\chi^2$
values up to $\sim 1.52$.  Statistically, the two models corresponding
to $\chi^2$ values of 0.83, and 0.84 are equally as good while the other
models in the subset with larger values of $\chi^2$ are poorer fits to the
interferometric data. Nevertheless, to be complete we computed profiles
for every model in the subset.  The equivalent widths for the two models
corresponding to the minimum  $\chi^2$ values had reasonable line shapes
compared to the observed profile but the model corresponding to $n = 2.2$
had an equivalent width of -25.0 \AA $\:$ while the model corresponding to $n=
2.1$ had an equivalent width of -21.6 \AA $\:$ which is closer to the value
for the observed spectra of -15.3 \AA.

\begin{figure}
\epsscale{.80}
\plotone{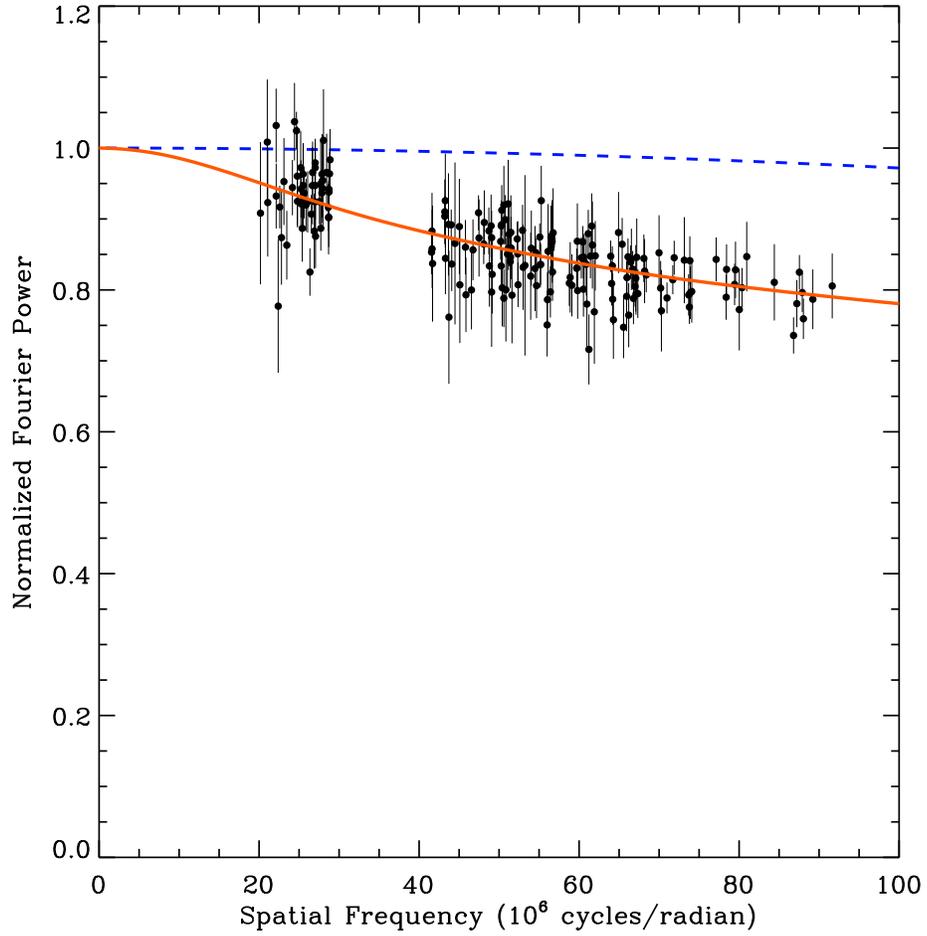}
\caption{The interferometric data and model with
$n=2.1 $, and $\rho_o = 3.0 \times 10^{-12}$ g/cm$^3$ for $\beta$ Psc. 
The dashed line 
represents the
central star and the solid line is obtained by taking a Fourier transform
of a synthetic image of the preferred model.
\label{fits_psc}}
\end{figure}

Figure~\ref{halpha_psc} shows the H$\alpha$ profile for the best-fit model
($n = 2.1$, $\rho_o = 3.0 \times 10^{-12}$ g/cm$^3$) in the left-most
panel in terms of interferometric observations and spectroscopy. Models
corresponding to other values of $n$ did not produce H$\alpha$
profiles that agreed with observations. Figure~\ref{halpha_psc} shows
the behavior of the H$\alpha$ profile with increasing values of $n$.
The reader can clearly see that the wings of the line in the middle and
right-most panels are too wide, resulting in predicted emission which is
too large.  These profiles correspond to models with $n =3.6$, $\rho_o =
1.5 \times 10^{-10}$ g/cm$^3$, equivalent width of -30.4 \AA, and  $n =4.0$,
$\rho_o = 4.0 \times 10^{-10}$ g/cm$^3$, equivalent width of -23.4 \AA, and
have interferometric reduced $\chi^2$ values of 1.13, and 1.21, respectively. 
Models with values
of $n < 2.1$ had large $\chi^2$ values based on
the interferometric results, and were eliminated.
Figure~\ref{fits_psc} shows the interferometric observations for
our best fit model with  $n = 2.1$, and $\rho_o =  3 \times 10^{-12}$ g/cm$^3$ 
for $\beta$ Psc.

\subsection{$\upsilon$ Cygnus} 

$\upsilon$ Cyg has long been recognized as a Be star with strong H$\alpha$
emission \citep{fle91, cam95}. Although several authors have noted that
this emission stays relatively constant over periods as long as decades
\citep[see, for example,][]{pet79} others have noted periodic outbursts in
H$\alpha$ emission, typical of many Be stars \citep{bal95, nei05}. The
interferometric observations for $\upsilon$ Cygnus are from 12 nights
between 28 Aug.\ 2005 and 26 Sept.\ 2005, and the H$\alpha$ line profile 
was acquired on
16 Sept.\ 2005 (see Table~\ref{obs_table}).  We note that we also have 
observations 
from 17 Oct.\ 2005 that confirm the spectroscopic stability of the star.

\citet{nei05} suggested that the outbursts, which occur on periods
of $\sim$ years, are the result of multi-periodic, non-radial pulsations.
\citet{nei05} derived the stellar parameters for this star by detailed
modeling constrained by photometric and spectroscopic data.  Three of
their models include veiling effects, and gravitational darkening for
a range of stellar rotational velocities, 0.80, 0.90, and 0.95 of the
critical velocity. (See \citet{nei05} for more detail.)  We have
chosen to adopt the stellar parameters from their model~D which has an
intermediate value of the rotational velocity (that is, 0.90 of critical
velocity).  Also, \citet{nei05} finds a value for the inclination angle
of $i\, \sim27^o$.
The stellar radius listed in Table~\ref{stellar_param}
corresponds to the equatorial radius given by \citet{nei05}. This is
a reasonable choice since the star is nearly pole-on.  The stellar
parameters adopted for $\upsilon$ Cyg are reasonable compared to other
values presented in the literature of this spectral type \citep[see
for example][]{cox00}.

The procedure for selecting the grid of models to be compared with
interferometric observations was obtained by the same selection process
as done for the two other stars in this investigation.  The final
selection of parameters to be searched ranged from $n = 1.8$ to $n =
4.5$, and $\rho_o = 1.0 \times 10^{-12}$ g/cm$^3$ to $\rho_o = 4.0 \times
10^{-9}$g/cm$^3$, giving a total of 193 models constructed for comparison
with interferometric observations.  The corresponding $\chi^2$ values
had a minimum value of 1.14 corresponding to the model with $n=4.0$ and
$\rho_o = 8.0 \times 10^{-10}$ g/cm$^3$ and a maximum value of 471
corresponding the model with $n=1.8$ and $\rho_o = 8.0 \times 10^{-12}$
g/cm$^3$.  The $\chi^2$ values revealed several models that statistically
fit the observations equally well. Again, we turn to spectroscopy to remove
the degeneracy.

\begin{figure}
\epsscale{.80}
\plotone{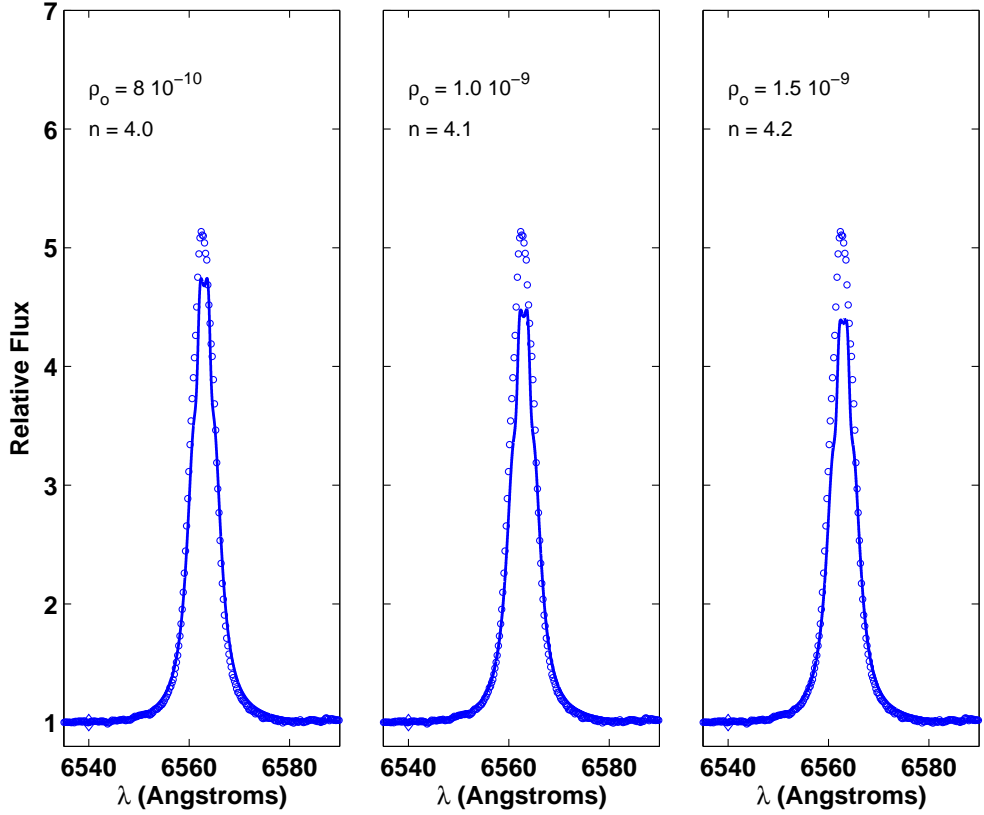}
\caption{The observed H$\alpha$ line for $\upsilon$ Cyg compared with the 
best predicted lines from the models which are also in agreement with 
the interferometric 
observations. The model profiles are the solid lines and the circles 
represent the observed line.  See Table~\ref{cyg_best} for the model 
parameters, H$\alpha$
equivalent widths, and the reduced $\chi^2$ values from 
interferometry corresponding to these models.
\label{halpha_cyg}}
\end{figure}

\begin{figure}
\epsscale{.80}
\plotone{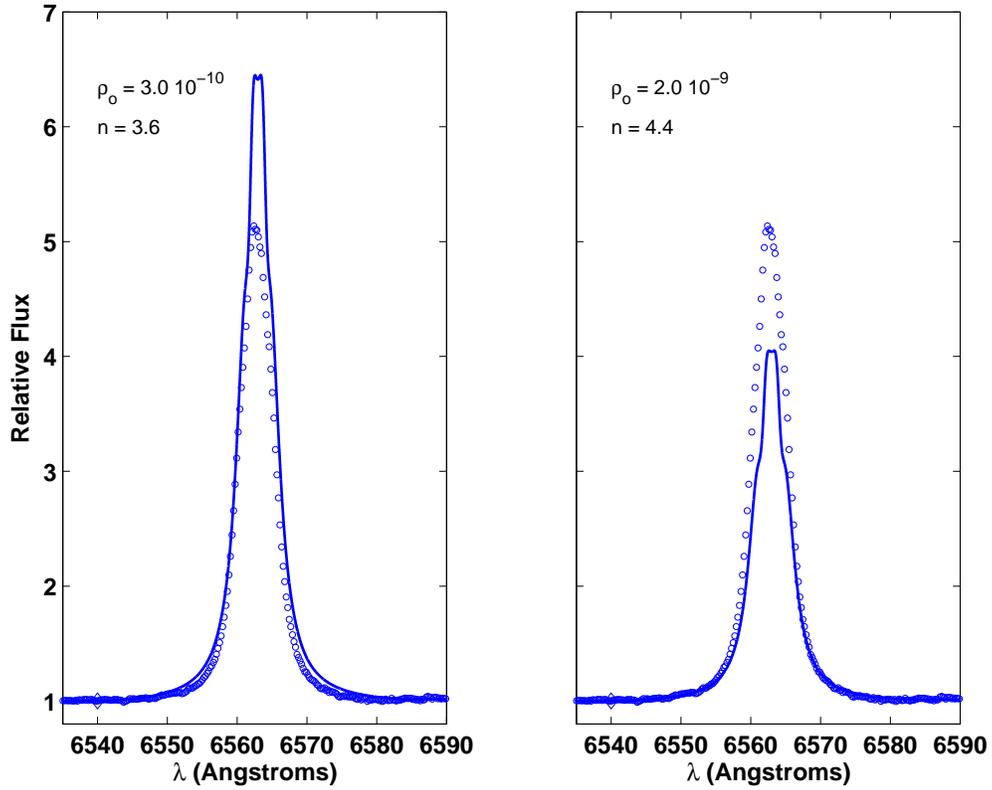}
\caption{The observed H$\alpha$ lines for $\upsilon$ Cyg compared with the 
best predicted lines for $n = 3.6$ and $n= 4.4$.  The model profiles are the 
solid lines and the circles 
represent the observed line.  The model on the left panel 
has a lower value of $n$ and higher density, resulting in an H$\alpha$ profile 
which is too large.  The model on the right panel 
has a higher value of $n$ and lower density, resulting in an H$\alpha$ profile 
which is too small.  
\label{halpha_cyg_bad}}
\end{figure}

\begin{figure}
\epsscale{.80}
\plotone{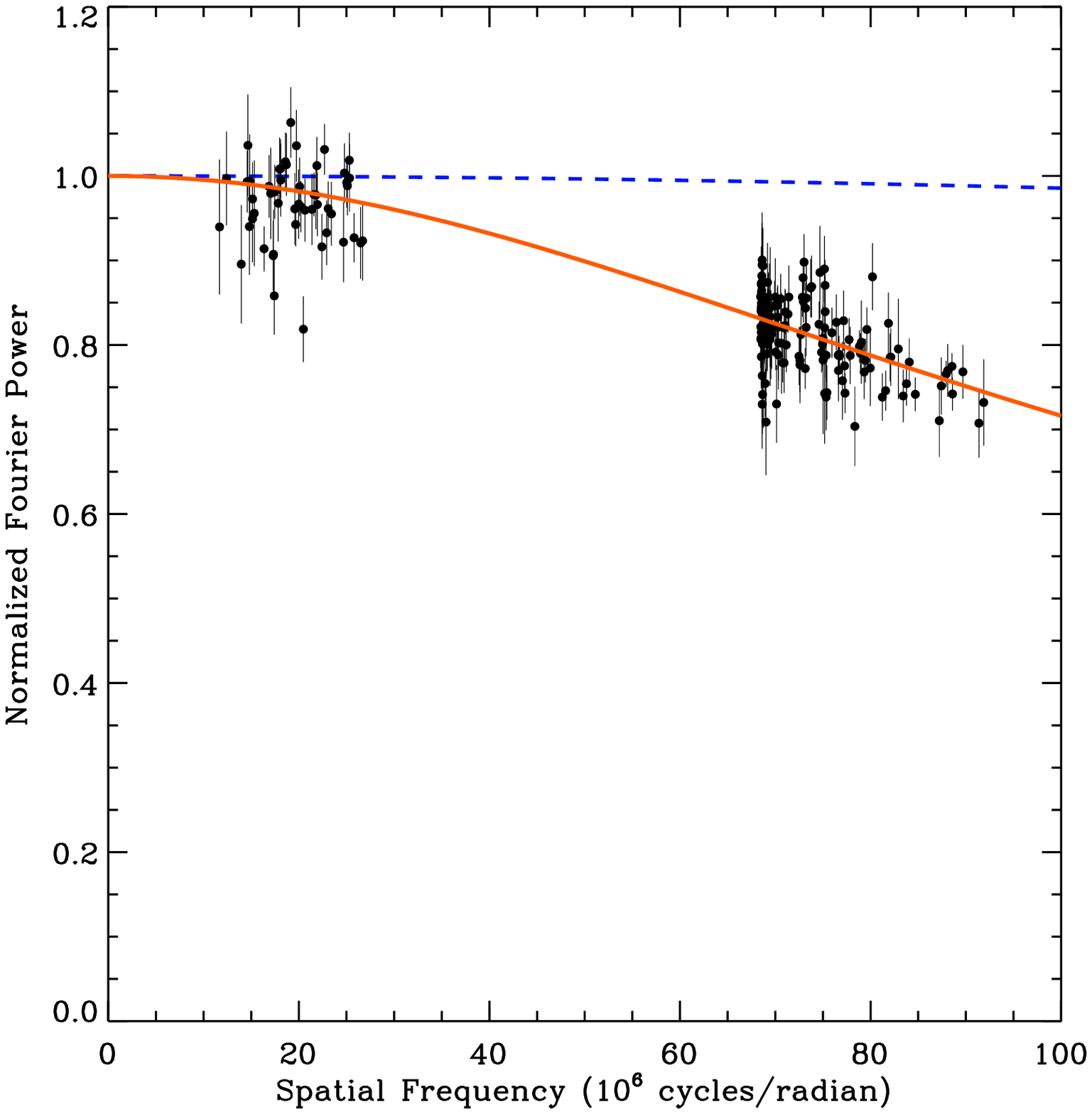}
\caption{ The interferometric data and model 
with $n=4.0$, and
$\rho_o = 8.0 \times 10^{-10}$ g/cm$^3$ for $\upsilon$ Cyg. The dashed line 
represents the
central star and the solid line is obtained by taking a Fourier transform
of a synthetic image of the preferred model.
\label{cyg_fits_interfer}}
\end{figure}

In order to predict H$\alpha$ profiles we adopt $i = 30^{\rm o}$ which is
consistent with \citet{nei05}.  We calculate the H$\alpha$ equivalent
widths for a subset ($\sim 22$ models) chosen based on the interferometric
reduced $\chi^2 < 2 $.
Statistically,
the models with larger values of $\chi^2$ are poorer fits to
the interferometric data. The equivalent widths for the
subset ranged from a minimum of -19.0 \AA $\:$ to a maximum of -43.1 \AA. The
observed H$\alpha$ line has an equivalent width of -24.8 \AA, and there
were three models which had reasonable line shapes and equivalent widths
that agreed within $\pm 2$ \AA $\,$ and had statistically 
similar interferometric $\chi^2$ values.
Figure~\ref{halpha_cyg} shows the H$\alpha$ profiles corresponding to
these models. Table~\ref{cyg_best} lists the values of $n$, $\rho_o$,
H$\alpha$ equivalent width, and the reduced $\chi^2$ values based on 
comparisons from interferometry for
these 3 models.  The models with $n = 4.1$ and $n = 4.2 $ 
in Figure~\ref{halpha_cyg}
have predicted profiles that are too narrow and also do not fit the
peak of the observed profile. Since the $\chi^2$ values are statistically
similar, we adopt the model corresponding to $n= 4.0$ and 
$\rho_o = 8.0 \times 10^{-10}$  g/cm$^3$ 
in Figure~\ref{halpha_cyg} as the best-fit.
As a further illustration, Figure~\ref{halpha_cyg_bad} shows H$\alpha$
profiles corresponding to models with smaller and larger values of $n$ than 
those given in 
Table~\ref{cyg_best}.  Specifically, these profiles correspond to models with 
$n=3.6$, $\rho_o = 3.0 \times 10^{-10}$  g/cm$^3$
and  $n=4.4$, $\rho_o = 2.0 \times 10^{-9}$  g/cm$^3$ 
with reduced $\chi^2$ values of 1.37, and 1.29, respectively. 
The shapes of these profiles do not match the observed line.  The predicted
profile corresponding to $n=3.6$ has too much emission (with an equivalent 
width of -33.7 \AA) and is too wide
in the line wings. The predicted
profile corresponding to $n=4.4$, has too little emission (with an equivalent 
width of -20.3 \AA) and much of the profile is narrower than the observation.
Figure~\ref{cyg_fits_interfer} shows a comparison of the interferometric 
observations and the best-fit model for $\upsilon$~Cyg. 

\begin{table}
\begin{center}
\caption{Best-Fit Models for $\upsilon$ Cygnus} \label{cyg_best}
\begin{tabular}{lccr} 
\tableline\tableline
$n$& $\rho_o$& H$\alpha$ Equivalent& $\chi^2$ \tablenotemark{a}\\
&g/cm$^3$&Width [\AA]&\\
\tableline
4.0& $8.0 \times 10^{-10}$&-26.0&1.14\\
4.1&$1.0 \times 10^{-9}$&-24.4&1.16\\
4.2&$1.5 \times 10^{-9}$&-23.7&1.30\\
\tableline
\tablenotetext{a}{Based on comparison with interferometry.}
\end{tabular}
\end{center}
\end{table}

\section{Conclusions}

In this work, disk density models were obtained for the classical Be
stars $\kappa\;$Dra, $\beta\;$Psc, and $\upsilon\;$Cyg by matching the
observed
interferometric H$\alpha$ visibilities with Fourier transforms of 
synthetic images produced by theoretical models. It was
demonstrated that
the additional constraint obtained by the (co-temporal)
observed
H$\alpha$ line profile was critical to selecting a density
model. This was particularly evident in the case of $\kappa$ Dra. 
We have demonstrated that in the case of these 3 stars, 
only 
by additional observational constraints can the degeneracy in the 
interferometric best-fitting models be removed.

The best fit models to both the interferometric observations and the
observed H$\alpha$ line profiles for the three stars are summarized in
Table~\ref{summary}.
The range of $n$
is consistent with previous determinations based on other diagnostics
(such as IR excess see -- \citet{wat86}).  We note that the range 
of ``base" densities, $\rho_o$,
varies by over two orders of magnitude between the three stars.

\begin{table}
\begin{center}
\caption{Summary of Best-Fit Models} \label{summary}
\begin{tabular}{lcr} 
\tableline\tableline
Star&$n$& $\rho_o$\\
&&g/cm$^3$\\
\tableline
$\kappa\;$Dra&4.2& $1.5 \times 10^{-10}$\\
$\beta\;$Psc&2.1&$3.0 \times 10^{-12}$\\
$\upsilon\;$Cyg&4.0&$8.0 \times 10^{-10}$\\
\tableline
\end{tabular}
\end{center}
\end{table}

\acknowledgements
 
The Navy Prototype Optical Interferometer is a joint project of the
Naval Research Laboratory and the US Naval Observatory, in cooperation
with Lowell Observatory, and is funded by the Office of Naval Research
and the Oceanographer of the Navy.
We thank the Lowell Observatory for the telescope time used to obtain the 
H$\alpha$ line spectra presented in this paper.
This research was supported in part by NSERC,   
the Natural Sciences and Engineering Research Council of Canada. 
We thank L. Thomson for contributions to this work as an NSERC USRA in 2006.
{\it Facilities:} \facility{LO:42in()}, \facility{NPOI ()}

\end{document}